# Complementary Relationships between Entanglement and Measurement

Michael Steiner[1] and Ronald Rendell[2]
[1,2]Inspire Institute, Alexandria VA.

**Abstract**

Complementary relationships exist regarding interference properties of particles such as pattern visibility, predictability and distinguishability. Additionally, relationships are known between information gain $G$ and measurement disturbance $F$ for entangled spin pairs. The question of whether a similar complementary relationship between entanglement and measurement occurs is examined herein. For qubit systems, both measurement on a single system and measurements on a bipartite system are considered in regards to the entanglement. It is proven that $\bar{E} + D \leq 1$ holds where $\bar{E}$ is the average entanglement after a measurement is made and for which $D$ is a measure of the measurement disturbance of a single measurement. For measurements on a bipartite system shared by Alice and Bob, it is shown that $\bar{E} + \bar{G} \leq 1$, where $\bar{G}$ is the maximum average information gain regarding Alice's result that can be obtained by Bob. These results are generalized for arbitrary initial mixed states and as well to non-Hermitian operators. In the case of maximally entangled initial states, it is found that $D \leq E_L$ and $\bar{G} \leq E_L$ where $E_L$ is the entanglement loss due to measurement by Alice. We conclude that the amount of disturbance and average information gain that one can gain are strictly limited by entanglement.

## 1. Introduction

Combining the disparate information from separate measurements is what allows a deterministic causal description of a system to be possible in classical physics. However, in quantum physics, the results from different types of measurements cannot always be combined and a quantum physical phenomenon found by observing the same system with different experimental arrangements can be mutually exclusive. The separation between the observer and observed system in realizing measurements of events may be arranged in many ways corresponding to different conditions of observations and type of apparatus determining the particular aspect of the phenomenon we wish to observe. At the quantum level, the deterministic chain of events in classical physics instead becomes lines of similar possibilities, each weighed by an amplitude for probability of occurrence and closed by the irreversible click of a detector. In his 1927 lecture at the Volta Congress in Como, Italy [1], Bohr called this logical exclusion of phenomena from different experimental arrangements complementarity. Bohr's initial attempts at justifying the complementarity picture used the Heisenberg uncertainty relations and arguments in terms of disturbances to the system occurring during measurement.

However, depending on the particulars of the experiment, complementarity has more generally appeared to be enforced by a variety of other signatures for quantum behavior that have since been identified within quantum mechanics: entanglement, uncertainty, measurement disturbance, which-way information, visibility, and distinguishability, among others. The role of these various signatures in quantum interference experiments have turned out to be neither completely logically independent nor logical consequences of one another, recently showing that there are numerous ways of dissecting complementarity [2][3]. Bohr's concept of complementarity continues to play a role today with studies of interference involving experimental measurement techniques [4]

A variety of quantitative expressions of wave-particle complementarity relationships have been previously derived that weight which-way information against fringe visibility and other quantifiers in interferometric settings. For example, Greenberger et al derived complementary single-particle duality relationships $V^2 + P^2 \leq 1$ between predictability $P = |\rho_{11} - \rho_{22}|$ and visibility of the interference pattern $V = 2|\rho_{12}|$ for a particle within an interferometer [5]. Englert found that including detectors in the interferometer paths leads to a duality between path distinguishability and visibility $D^2 + V^2 \leq 1$ [6]. It has also been shown that in order to obtain information of a single qubit state, disturbance of the qubit is necessary. Busch showed in [7] that there is a limitation on quantum measurement for which information cannot be obtained without disturbance. Similarly, inequalities have been found between measurement sharpness and disturbance [8] and as well information gain versus state disturbance was reported in [9]. Experimental confirmation of various complementarity relations has been achieved for single quantum objects of increasingly larger size, approaching the mesoscopic and macroscopic levels [10][11][12][13][14][15].

Although these results show that for a single qubit various quantities are complementary, the issue of the potential complementary relationships between the entanglement of two qubits and measurement has not been reported. In order to examine potential complementary relationships between entanglement and measurement, let us consider a measurement that is made by Alice on one of two qubits $A$ that is initially entangled with a second qubit $B$ that is held by Bob. In [16] the authors showed simulation results for which entanglement of two parties is reduced when one of the particles is subject to a measurement via a particular measurement device model proposed by Gurvitz [17]. In this paper, such loss of entanglement under measurement is shown theoretically to be fundamental across a general class of positive operator valued measurements (POVM) i.e. not limited to any particular measurement device model.

A property of an entanglement measure or monotone is that the average entanglement that remains after a measurement on a subsystem is made is less than or equal to the initial entanglement [18]. Typically, the amount of average entanglement that remains after the measurement will depend on the disturbance $D$ of Alice's measurement which is closely related to the strength of her measurement; full loss of average entanglement can occur in the case of strong measurement with significant disturbance and no loss of entanglement in the limit of weak measurement with no substantial disturbance. We first examine the degree to which the average entanglement $\bar{E}$ remaining after Alice's measurement is related to the measurement disturbance $D$. Over a general class of POVMs that Alice can apply, it is proven that $\bar{E} + D \leq 1$ between average entanglement $\bar{E}$ and measurement disturbance $D$. Hence to the extent that Alice's measurement is strong or has a large disturbance, the entanglement that remains must be sufficiently small and visa-versa.

In the prior analysis, a measurement by Alice was made on a single qubit that was initially entangled with an ancilla. Additionally, it is desirable to understand the effect that Alice's measurement has on the amount of information that can be gained regarding Alice's outcome by a second measurement on the ancilla $B$ by a second experimenter, Bob. It has been shown previously that by varying the strength of a measurement, complementary trade-offs between information gain $G$ and measurement disturbance $F$ for entangled spin pairs have been found for which $F^2 + G^2 \leq 1$ [9, 19–22]. For bipartite systems, this is found to extend to a triality relating interference properties and entanglement, $P^2 + V^2 + C^2 \leq 1$ where $C$ is the concurrence measure of entanglement [23]. However, it is desirable to understand the extent to which the remaining average entanglement $\bar{E}$ limits information gain. One might expect that the existence of average entanglement $\bar{E}$ should limit the amount of information that Bob can obtain regarding Alice's measurement result. It is indeed the case; it is proven for arbitrary initial states that the average entanglement $\bar{E}$ and average information gain $\bar{G}$ are complementary in the sense that $\bar{E} + \bar{G} \leq 1$.

In the case of maximally entangled initial states a direct relationship is found between the change in entanglement such that $D \leq \bar{E}_L$ and $\bar{G} \leq \bar{E}_L$ where $\bar{E}_L \equiv -\Delta\bar{E}$ is the entanglement loss between the initial entanglement of unity and the entanglement after Alice's measurement. Although useful as a bound for maximally entangled states with no initial classical correlation, it is discussed in Sec. 2.1 why this result does not generalize to arbitrary initial states.

The paper is organized as follows. In Section 2.1, further background on the complementary aspects of entanglement and measurement are provided and as to why average entanglement should limit disturbance and information gain. Discussion is provided for the form of complementary relationships between entanglement and measurement. In the remainder of Section 2, measurement operations used throughout the paper as well as the entanglement measure $E$, measurement disturbance $D$, and information gain $G$ are also defined. Results in Section 3 are presented assuming two entangled particles. In Sec. 3.1 Alice makes a measurement on her qubit which has an associated measurement disturbance $D$. The effect of the entanglement is examined relative to the measurement disturbance. In Sec. 3.2 two-particle measurements of both Alice and Bob are considered and a tradeoff between entanglement and average information gain is presented. Results are extended to include all mixed initial states in Appendix 2 and to non-Hermitian operations in Appendix 3, found in the Supplementary Material. Conclusions are presented in Section 4.

## 2. Background

### 2.1. Complementary aspects of entanglement and measurement

Schrödinger [24] had noted as early as 1935 that entanglement can result in the interaction of two systems for which, *"after a time of mutual influence the systems separate again, then they can no longer be described the same way as before, viz., by endowing each of them with a representative state of its own"* and that this is a characteristic trait of quantum theory, *"I would not call that one but the characteristic trait of quantum mechanics, the one that enforces its entire departure from classical lines of thought."* It was further found by Schrödinger that the particle cannot be said to be in any particular pure state, but must be described by a density matrix [25][26]. The fact that an overall maximally entangled system cannot be described by a product state of individual subsystems and for which there is an observable violation for such systems was later formalized by Bell's inequality and the subsequent verification of the prediction of entanglement via the experimental violation of Bell's inequality.

To further illustrate the rationale for situations whereby entanglement and measurement can be considered complementary, consider as in Sec. 1 that Alice and Bob have Qubits $A$ and $B$ respectively that are in an initially maximally entangled state. Each qubit is in a maximally mixed state and lacks a well-defined pure state, as first noted by Schrödinger [25]. In the case that Alice's measurement is very weak, Alice's qubit would be expected to remain on-average nearly fully entangled and cannot be in a well-defined pure state that is correlated with Alice's result. Hence one would expect that a second measurement by Bob on Qubit $B$ cannot provide any significant information as to what Alice measured due to the entanglement that still exists between Qubits $A$ and $B$. On the other hand, suppose that Alice's measurement is a strong measurement and completely destroys the entanglement that initially existed. After Alice's measurement, Qubits $A$ and $B$ are both projected into well-defined correlated pure states, allowing Bob to obtain information regarding Alice's result by measuring Qubit B. In such a case, the amount of average entanglement that remains after Alice's measurement would be expected to limit the amount of information regarding Alice's outcome that can be obtained by Bob. This is because to the extent that average entanglement remains after Alice's measurement, such entanglement is representative of the extent that the subsystems remain in a mixed state. Such a mixed state cannot be correlated with Alice's measurement results and hence strictly limits the ability of Bob to obtain information regarding Alice's outcome for which it can be found that the complementary relationship $\bar{E} + \bar{G} \leq 1$ exists between the average entanglement $\bar{E}$ and average information gain $\bar{G}$.

In the relationship above, the average information gain $\bar{G}$ that Bob can gain regarding Alice's outcome is found to be complementary to the average entanglement $\bar{E}$ after Alice's measurement. One might consider the possibility of deriving complementary relationships between the information that Bob can gain regarding Alice's measurement and the initial entanglement $E$ of Qubits A and B *prior* to a measurement by Alice and Bob. However, it does not appear that the initial entanglement $E$ by itself (without further modification) is generally complementary to the information gain. Consider a maximally entangled initial state for which the initial entanglement is $E = 1$. In this case, a measurement conducted by both Alice and Bob in the same basis for which Alice obtains 0 (or 1) will be followed by Bob obtaining Alice's result 0 (or 1) resulting in $G = 1$ information gain that Bob can gain regarding Alice's outcome. Hence for such an initial state we have $E = 1, G = 1$. Now, consider an initial mixed joint Bob-Alice density matrix $\rho_{BA}$ qubit state as $\rho_{BA} = \frac{1}{2}|00\rangle\langle 00| + \frac{1}{2}|11\rangle\langle 11|$. Such an initial density matrix has classical correlation in the following sense. A measurement conducted by both Alice and Bob in the computational basis for which Alice obtains 0 (or 1) will be followed by Bob obtaining Alice's result 0 (or 1). For such an initial state, the information gain $G$ that Bob can gain regarding Alice's outcome is $G = 1$, and for this case we have $E = 0, G = 1$. Hence depending on the initial state, the information gain remains the same i.e. $G = 1$ yet the entanglement is seen to vary from its potential minimum of $E = 0$ to its potential maximum of $E = 1$. This counterexample of $\rho_{BA} = \frac{1}{2}|00\rangle\langle 00| + \frac{1}{2}|11\rangle\langle 11|$ illustrates that the initial entanglement and information gain are not complementary in-general.

One can obtain an additional inequality as a lemma to the relationship $\bar{E} + \bar{G} \leq 1$ in the case when one restricts the initial state to a maximally entangled initial state. Starting with $\bar{E} + \bar{G} \leq 1$, this can be rewritten as $\bar{E} + \bar{G} \leq E_i$, where $E_i$ is the initial entanglement of Qubits A and B. Consider that after Alice's measurement the change in entanglement from the initial to the final is given by $\Delta \bar{E} = \bar{E} - E_i$. Upon defining the entanglement loss $\bar{E}_L \equiv -\Delta \bar{E}$ between the initial entanglement and final average entanglement and the entanglement after Alice's measurement, we have $\bar{G} \leq \bar{E}_L$. Hence a direct relationship can be found $\bar{G} \leq \bar{E}_L$ (and by a similar argument $D \leq \bar{E}_L$) for which the information gain must be less than the loss of entanglement. That is, in order to gain information when starting with a maximally entangled state, there must be some corresponding loss of entanglement. Now, from the discussion above regarding the counterexample of initial classical states such as $\rho_{BA} = \frac{1}{2}|00\rangle\langle 00| + \frac{1}{2}|11\rangle\langle 11|$ one can see why such a relationship makes sense. By restricting the initial state to a maximally entangled state, the initial state is completely pure and such a quantum state appears to be devoid of its mixed classical counterpart which also conveys some classical information to Bob regarding Alice's outcome. Hence, it is found that the entanglement loss is indeed directly complementary to both measurement disturbance and information gain for initial maximally entangled states, which are widely utilized in both theory and experiment in quantum information.

2.2. *Entanglement quantification*

Consider a two-qubit system in an arbitrary pure state given by

$$|\psi_0\rangle = a_{00}|00\rangle + a_{01}|01\rangle + a_{10}|10\rangle + a_{11}|11\rangle \tag{1}$$

This can be rewritten via a Schmidt decomposition as

$$|\psi_0\rangle = \sqrt{\alpha_0}|u_1\rangle \otimes |v_1\rangle + \sqrt{1-\alpha_0}|u_2\rangle \otimes |v_2\rangle. \tag{2}$$

where $|u_i\rangle \in \mathcal{H}_1$ and $|v_i\rangle \in \mathcal{H}_2$, $i = 1, 2$ and $\mathcal{H}_1, \mathcal{H}_2$ are the Hilbert spaces of qubit 1 and 2 respectively, and we can assume without loss of generality that $\alpha_0 \leq .5$. Additionally, the $|u_i\rangle$ are chosen orthogonally in $\mathcal{H}_1$ and similarly for the $|v_i\rangle$ in $\mathcal{H}_2$. Note that $|\psi_0\rangle$ is of the form of the two-qubit ancilla state Eq. (2) when $|u_1\rangle = |0\rangle, |v_1\rangle = |0\rangle, |u_2\rangle = |1\rangle, |v_2\rangle = |1\rangle$.

In order to quantify the entanglement throughout the paper, note that it is known that all entanglement measures of pure states are equivalent in the sense that a one-to-one relationship can be found between any two entanglement measures [27]. The well-known entropy of entanglement which we denote as $E_H$ is found as the von Neumann entropy of either reduced density matrix of a bipartite system. The von Neumann entropy is a function of the two eigenvalues of either reduced density matrix $\{\lambda, 1-\lambda\}$ given in bits by

$$E_H(\lambda) = -\lambda \log_2 \lambda - (1-\lambda)\log_2(1-\lambda). \tag{3}$$

Another measure that is equivalent (in the sense given in [27]) to the von Neumann entropy is twice the minimum eigenvalue of the reduced density matrix which will be denoted $E(\lambda_{\min})$. This is also equivalent to the "geometric measure of entanglement" [28]. Since $\lambda_{\min} + \lambda_{\max} = 1$, when $\lambda \leq 1/2$ then

$$E(\lambda) = 2\lambda \tag{4}$$

which we have chosen for our entanglement measure and we will demonstrate that a direct relationship with measurement will be established using this measure. For a state given by the Schmidt decomposition of Eq. (2) one also finds a direct relationship with the smallest Schmidt coefficient $\alpha_0$ for which $E(\lambda) = 2\alpha_0$.

2.3. *Measurement operations*

Let us consider a protocol in which Bob and Alice each has a qubit in the initial pure state $|\psi_0\rangle \in \mathcal{H}_1 \otimes \mathcal{H}_2$ represented by the Schmidt decomposition of Eq. (2) for which $\alpha_0 \leq .5$ and the $|u_i\rangle$ are chosen orthogonally in $\mathcal{H}_1$ and similarly the $|v_i\rangle$ in $\mathcal{H}_2$. We define projection onto a pure state as a strong measurement (a sharp measurement is defined in Busch [7] as a projection-valued measurement which can include higher rank projections). Alice makes a measurement with two possible outcomes {0,1} on the qubit defined on $\mathcal{H}_2$. A two-outcome positive operator valued measurement (POVM) that Alice can apply is found as

$$M_0^{(A)} \equiv \frac{1}{\sqrt{1+2\Lambda(1+\Lambda)}} \begin{bmatrix} \Lambda + \cos^2\theta & e^{i\phi}\sin\theta\cos\theta \\ e^{-i\phi}\sin\theta\cos\theta & \Lambda + \sin^2\theta \end{bmatrix}$$

$$M_1^{(A)} \equiv \frac{1}{\sqrt{1+2\Lambda(1+\Lambda)}} \begin{bmatrix} \Lambda + \sin^2\theta & -e^{i\phi}\sin\theta\cos\theta \\ -e^{-i\phi}\sin\theta\cos\theta & \Lambda + \cos^2\theta \end{bmatrix} \quad (5)$$

Where $0 \leq \phi \leq 2\pi$, $0 \leq \Lambda$, $\theta, \phi, \Lambda \in \mathbb{R}$, where $\mathbb{R}$ denotes the real numbers and $M_0^{(A)\dagger} M_0^{(A)} + M_1^{(A)\dagger} M_1^{(A)} = I$. Note that any set of matrices $M_i^{(A)} \in S^2$, $i = 0,1$ where $S^2$ denotes the set of $2 \times 2$ positive semi-definite matrices that also satisfies $M_0^{(A)\dagger} M_0^{(A)} + M_1^{(A)\dagger} M_1^{(A)} = I$ is described by the class defined by Eq. (5). We restrict $\Lambda + \cos^2\theta \geq \Lambda + \sin^2\theta$ so that $M_0^{(A)}$ projects with a higher probability into state $|0\rangle$ than $|1\rangle$ and visa-versa for $M_1^{(A)}$; this is achieved when $0 \leq \theta \leq \frac{\pi}{4}$. When outcome $i \in \{0,1\}$ occurs corresponding to $M_i^{(A)}$, the partial density matrices of Alice and Bob are denoted $\rho_{A,i}$ and $\rho_{B,i}$ respectively, and where $p_{A,i}$ is the probability of Alice obtaining result $M_i^{(A)}$, $i = 0,1$, $p_{A,i} = \text{Tr}(M_i^{(A)\dagger} M_i^{(A)} |\psi_0\rangle\langle\psi_0|)$.

## 2.4. Measurement disturbance and information gain

The effects of measurement will be quantified considered using several quantities assuming an initially entangled state shared by Alice and Bob. [20–23] In the case that Alice makes a single measurement, the disturbance $D$ will be used to quantify the effect of the measurement in Sec. 2.4.1. In the case that Alice and Bob both make measurements on the respective qubits of an entangled state, a bipartite measure of information gain $G$ will be defined in Sec. 2.4.2. [20–23]

### 2.4.1. Quantification of measurement disturbance

In the case of two entangled particles that Alice and Bob share and a measurement is made by Alice on her qubit, the effect of the measurement will be quantified by the use of the measurement disturbance or quality factor $F$ as utilized in [19–22]. In order to examine the tradeoff between entanglement and measurement, we want to determine the extent that if there is measurement, the initial entanglement is affected. For our purposes, we define the measurement disturbance as $D = 1 - F$ which was utilized in [19] and is shown in Appendix 4 as $D = \frac{1}{1+2\Lambda(1+\Lambda)}$. Note that the disturbance is not a function of the degree of superposition of the qubit, nor the parameters $\theta$ and $\phi$ in Eq. (5), but rather it only enters through the strength or weakness of the measurement which is a function of $\Lambda$.

### 2.4.2. Information gain in two-particle measurements

A second quantity that was utilized in quantifying the effect of measurement in [19–22] is the information gain. This will be presented and extended for the use of two-particle measurements which will be applied in Sec. 3.2. For single qubit weak measurement in both [19, 21], the information gain is defined by

$$G = 1 - |\langle 1|\psi_{P,\uparrow}\rangle|^2 - |\langle 0|\psi_{P,\downarrow}\rangle|^2. \tag{6}$$

As noted in [19][21] in addition to showing that $G$ represents the precision of the measurement, Eq. (6) represents a probability of error. In the case of [21], an assumption for simplicity is made that there is symmetric ambiguousness or that $|\langle 1|\psi_{P,\uparrow}\rangle|^2 = |\langle 0|\psi_{P,\downarrow}\rangle|^2$. Note that the probability of error whereby a spin $|\uparrow\rangle$ ($|\downarrow\rangle$) is measured by a result $|1\rangle$ ($|0\rangle$) is given by

$$P_e = |\langle 1|\psi_{P,\uparrow}\rangle|^2 \beta + |\langle 0|\psi_{P,\downarrow}\rangle|^2 (1-\beta) \tag{7}$$

where $\beta$ is probability of spin $|\uparrow\rangle$ occurring from Eq. (4.1) of Appendix 4 of the Supplementary Material. In the case that $|\langle 1|\psi_{P,\uparrow}\rangle|^2 = |\langle 0|\psi_{P,\downarrow}\rangle|^2 \equiv \alpha$, this simplifies to $P_e = \alpha$ or $G = 1 - 2\alpha$ or $G = 1 - 2P_e$. Note that since $0 \leq P_e \leq 1/2$ for optimal measurements, one can see that $0 \leq G \leq 1$. Let us define the following extension of the gain for single-particle measurement when $|\langle 1|\psi_{P,\uparrow}\rangle|^2 \neq |\langle 0|\psi_{P,\downarrow}\rangle|^2$ as $G = 1 - 2P_e$ or

$$G = 1 - 2|\langle 1|\psi_{P,\uparrow}\rangle|^2 \beta - 2|\langle 0|\psi_{P,\downarrow}\rangle|^2 (1-\beta) \tag{8}$$

which reduces to Eq. (6) when $|\langle 1|\psi_{P,\uparrow}\rangle|^2 = |\langle 0|\psi_{P,\downarrow}\rangle|^2$. This definition can be further generalized as follows. Suppose Alice and Bob share an arbitrary initial bipartite two-qubit state and Alice first makes a measurement in the basis $\{M_0^{(A)}, M_1^{(A)}\}$. Bob's task is to attempt to determine as best as possible in the sense of minimizing the probability of error, what Alice's measurement result was. Bob in-general will be assumed to know the operators $\{M_0^{(A)}, M_1^{(A)}\}$, but not Alice's result. In this case we will utilize the measure of information gain defined as $G = 1 - 2P_e$. Now, this measure of gain can be applied directly to the measurement operations specified previously. Let us define $G_i \equiv 1 - 2P_{e|A=i}$, where $P_{e|A=i}$ is the probability of error given Alice obtains result $i \in \{0,1\}$ and let us define $\pi_{A,i}$ as the probability of Alice obtaining result $i \in \{0,1\}$. The average information gain that Bob obtains about Alice's qubit is then found as $\bar{G} = \pi_{A,0} G_0 + \pi_{A,1} G_1$.

## 3. Results

We will now consider that Alice and Bob both share an entangled state. In the first case in Sec. 3.1, the effect of the entanglement when Alice makes a measurement with some disturbance $D$ defined in Sec. 2.4.1 will be considered while in Sec. 3.2 both Alice and Bob will make measurements with the goal of maximizing the information gain as defined in Sec. 2.4.2. In the case of two-qubit measurement, without loss of generality, the initial state can be taken to be of the form $\sqrt{\alpha}|0\rangle \otimes |0\rangle + \sqrt{1-\alpha}|1\rangle \otimes |1\rangle$ as shown in Appendix 1 of the Supplementary Material. Alice applies the measurement consisting of the operators $M_i^{(A)}$ of Eq. (5) for which the entanglement will be reduced from its original value of $2\alpha_i$ to one of two possible final values depending on the outcome of Alice's measurement $\{0,1\}$. The average entanglement that is obtained is given by

$$\bar{E} = \pi_{A,0} E(\rho_{B,0}) + \pi_{A,1} E(\rho_{B,1}). \tag{9}$$

where $\rho_{B,i}$ is the partial density matrix of Bob when outcome $i$ occurs (note that since $E(\rho_{B,i}) = E(\rho_{A,i})$ for bipartite systems, one can equivalently use Alice's partial density matrix to compute the entanglement). In the case of one-particle measurement, the measurement refers only to Alice's measurement and the goal is to determine the tradeoff between the measurement disturbance $D$ and the average entanglement $\bar{E}$.

In the case of two-particle measurement, Alice similarly applies the measurement consisting of the operators $M_i^{(A)}$ of Eq. (5) for which the average entanglement is obtained in Eq. (9). At this point, Bob makes a measurement, attempting to maximize his information gain regarding Alice's result. That is, he desires to make a measurement that will be maximally correlated with Alice's result so that his probability of error is minimized. Bob is assumed to know Alice's measurement basis. In order to optimize the probability of error, Bob must measure in a basis that is optimal in terms of minimizing the probability of error. A solution to Bob's basis problem is given by Bergou in [29]. Bergou's construction is utilized to optimize the information gain for two-particle measurement. Note that Alice and Bob can apply local unitary operations to their respective qubits without changing the entanglement.

### 3.1. Two entangled particles, single measurement by Alice

In this section, we will assume that Bob and Alice each have a particle of an entangled state and Alice desires to measure her qubit. As we will see, Alice's measurement will generally reduce the entanglement of the shared state of Alice and Bob to a final value that depends on the strength of the measurement. Alice will apply the general measurement operators $\{M_0^{(A)}, M_1^{(A)}\}$ of the form shown in Eq. (5). One might also desire for completeness to consider what effect the interaction of Alice's macroscopic measurement device has on the results. To this end, the von Neumann measurement scheme, which includes consideration of measurement device pointer states, is considered in Appendix 4. It is shown in Appendix 4 that particle interaction with a measurement device followed by projection within the measurement device is completely equivalent to Alice using a direct measurement via her operators $\{M_0^{(A)}, M_1^{(A)}\}$. Given an initial state and measurement operators of the form of Eq. (5), it is also found that $F = \frac{\Lambda}{\sqrt{1+2\Lambda(1+\Lambda)}}$ which is only a function of the strength or weakness $\Lambda$ of Alice's measurement operators. Due to the equivalence found in Appendix 4, it will henceforth be simply considered that Alice makes a measurement on her qubit via the operators $\{M_0^{(A)}, M_1^{(A)}\}$, without further reference to the von Neumann measurement technique.

The entanglement after Alice's initial measurement is given by Eq. (9). The probability of Alice obtaining result $i \in \{0,1\}$, $\pi_{A,i}$ and $E(\psi_{A,i})$ are now evaluated,

$$\pi_{A,i} = \text{Tr}\left[|\psi_0\rangle\langle\psi_0|(I\otimes M_i^{(A)})^\dagger(I\otimes M_i^{(A)})\right]. \tag{10}$$

Substituting for $|\psi_0\rangle$ and $M_i^{(A)}$ and simplifying,

$$\pi_{A,0} = \frac{1 + 2\Lambda(1+\Lambda) + (-1+2a)(1+2\Lambda)\text{Cos}[2\theta]}{2 + 4\Lambda(1+\Lambda)}$$

$$\pi_{A,1} = \frac{1}{2} - \frac{(-1+2a)(1+2\Lambda)\text{Cos}[2\theta]}{2+4\Lambda(1+\Lambda)}. \tag{11}$$

The entanglement depends on the particular outcome that Alice finds. The density matrix that Bob obtains in the case that outcome $i \in \{0,1\}$ occurs is found by

$$\rho_{B,i} = \frac{\text{Tr}_A\left[(I \otimes M_i^{(A)})|\psi_0\rangle\langle\psi_0|(I \otimes M_i^{(A)})^\dagger\right]}{\pi_{A,i}}. \tag{12}$$

where $\text{Tr}_A$ denotes the partial trace operation on Alice's Hilbert space. Substituting and reducing, it is found that:

$$\rho_{B,0} = \begin{bmatrix} \frac{2a(\Lambda^2 + (1+2\Lambda)\text{Cos}[\theta]^2)}{1+2\Lambda(1+\Lambda)+(-1+2a)(1+2\Lambda)\text{Cos}[2\theta]} & \frac{\sqrt{-((-1+a)a)}e^{-i\phi}(1+2\Lambda)\text{Sin}[2\theta]}{1+2\Lambda(1+\Lambda)+(-1+2a)(1+2\Lambda)\text{Cos}[2\theta]} \\ \frac{\sqrt{-((-1+a)a)}e^{i\phi}(1+2\Lambda)\text{Sin}[2\theta]}{1+2\Lambda(1+\Lambda)+(-1+2a)(1+2\Lambda)\text{Cos}[2\theta]} & \frac{(-1+a)(-1-2\Lambda(1+\Lambda)+(1+2\Lambda)\text{Cos}[2\theta])}{1+2\Lambda(1+\Lambda)+(-1+2a)(1+2\Lambda)\text{Cos}[2\theta]} \end{bmatrix}$$

$$\rho_{B,1} = \begin{bmatrix} \frac{a(-1-2\Lambda(1+\Lambda)+(1+2\Lambda)\text{Cos}[2\theta])}{-1-2\Lambda(1+\Lambda)+(-1+2a)(1+2\Lambda)\text{Cos}[2\theta]} & \frac{\sqrt{-((-1+a)a)}e^{-i\phi}(1+2\Lambda)\text{Sin}[2\theta]}{-1-2\Lambda(1+\Lambda)+(-1+2a)(1+2\Lambda)\text{Cos}[2\theta]} \\ \frac{\sqrt{-((-1+a)a)}e^{i\phi}(1+2\Lambda)\text{Sin}[2\theta]}{-1-2\Lambda(1+\Lambda)+(-1+2a)(1+2\Lambda)\text{Cos}[2\theta]} & \frac{(-1+a)(1+2\Lambda(1+\Lambda)+(1+2\Lambda)\text{Cos}[2\theta])}{-1-2\Lambda(1+\Lambda)+(-1+2a)(1+2\Lambda)\text{Cos}[2\theta]} \end{bmatrix}. \tag{13}$$

The entanglement $E(\rho_{B,i})$ is found as twice the minimum eigenvalue of the corresponding matrices in Eq. (13). These are found to simplify to

$$E(\rho_{B,0}) = 1 - \frac{2\sqrt{16(-1+a)a\Lambda^2(1+\Lambda)^2 + (1+2\Lambda(1+\Lambda)+(-1+2a)(1+2\Lambda)\text{Cos}[2\theta])^2}}{2+4\Lambda(1+\Lambda)+2(-1+2a)(1+2\Lambda)\text{Cos}[2\theta]}$$

$$E(\rho_{B,1}) = 1 - \frac{2\sqrt{16(-1+a)a\Lambda^2(1+\Lambda)^2 + (1+2\Lambda(1+\Lambda)-(-1+2a)(1+2\Lambda)\text{Cos}[2\theta])^2}}{2+4\Lambda(1+\Lambda)-2(-1+2a)(1+2\Lambda)\text{Cos}[2\theta]}. \tag{14}$$

The average entanglement can now be computed by substituting Eqs. (11) and (14) into Eq. (9). Upon simplifying, one obtains for $\bar{E}$

$$-\frac{1}{2+4\Lambda(1+\Lambda)}(-2-4\Lambda-4\Lambda^2+\sqrt{16(-1+a)a\Lambda^2(1+\Lambda)^2+(1+2\Lambda(1+\Lambda)-(-1+2a)(1+2\Lambda)\text{Cos}[2\theta])^2}$$
$$+\sqrt{16(-1+a)a\Lambda^2(1+\Lambda)^2+(1+2\Lambda(1+\Lambda)+(-1+2a)(1+2\Lambda)\text{Cos}[2\theta])^2}) \tag{15}$$

Now, consider the function $H \equiv \bar{E} + D - 1$. Note that both $\bar{E}$ and $D$ are independent of $\phi$, hence $H$ is independent of $\phi$. $H$ can be shown to simplify to

$$-\frac{1}{2+4\Lambda(1+\Lambda)}(-2+\sqrt{16(-1+a)a\Lambda^2(1+\Lambda)^2+(1+2\Lambda(1+\Lambda)-(-1+2a)(1+2\Lambda)\text{Cos}[2\theta])^2}$$
$$+\sqrt{16(-1+a)a\Lambda^2(1+\Lambda)^2+(1+2\Lambda(1+\Lambda)+(-1+2a)(1+2\Lambda)\text{Cos}[2\theta])^2}) \tag{16}$$

Since $\bar{E} + D \leq 1$ iff $H \leq 0$, and since $-\frac{1}{2+4\Lambda(1+\Lambda)} < 0$, we can multiply both sides by $-\frac{1}{2+4\Lambda(1+\Lambda)}$ and it is needed to be shown that $\hat{H}(a, \Lambda, \theta) \geq 0$, where

$$\hat{H}(a, \Lambda, \theta) =$$
$$-2+\sqrt{16(-1+a)a\Lambda^2(1+\Lambda)^2+(1+2\Lambda(1+\Lambda)-(-1+2a)(1+2\Lambda)\text{Cos}[2\theta])^2}$$
$$+\sqrt{16(-1+a)a\Lambda^2(1+\Lambda)^2+(1+2\Lambda(1+\Lambda)+(-1+2a)(1+2\Lambda)\text{Cos}[2\theta])^2}. \tag{17}$$

Now, $\hat{H}$ in Eq. (17) is a function of $a, \Lambda, \theta$. It will now be shown that $\arg\min_a \hat{H}(a, \Lambda, \theta)$ where $a \in \left[0, \frac{1}{2}\right]$ occurs at $a = 1/2$. Taking the derivative of $\hat{H}(a, \Lambda, \theta)$ with respect to $a$ it is found

$$\frac{\partial \hat{H}(a, \Lambda, \theta)}{\partial a}$$
$$= -\frac{1}{2+4\Lambda(1+\Lambda)}\left(\frac{2(4(-1+2a)\Lambda^2(1+\Lambda)^2-(1+4\Lambda+6\Lambda^2+4\Lambda^3)\text{Cos}[2\theta]+(-1+2a)(1+2\Lambda)^2\text{Cos}[2\theta]^2)}{\sqrt{16(-1+a)a\Lambda^2(1+\Lambda)^2+(1+2\Lambda(1+\Lambda)-(-1+2a)(1+2\Lambda)\text{Cos}[2\theta])^2}}\right.$$
$$\left.+\frac{2(4(-1+2a)\Lambda^2(1+\Lambda)^2+(1+4\Lambda+6\Lambda^2+4\Lambda^3)\text{Cos}[2\theta]+(-1+2a)(1+2\Lambda)^2\text{Cos}[2\theta]^2)}{\sqrt{16(-1+a)a\Lambda^2(1+\Lambda)^2+(1+2\Lambda(1+\Lambda)+(-1+2a)(1+2\Lambda)\text{Cos}[2\theta])^2}}\right) \tag{18}$$

We want to determine where the local minimal values of $\hat{H}(a, \Lambda, \theta)$ occur. Setting the above to zero, we can multiply through by $-(2 + 4\Lambda(1 + \Lambda))$, and this will occur when

$$\frac{2(4(-1+2a)\Lambda^2(1+\Lambda)^2-(1+4\Lambda+6\Lambda^2+4\Lambda^3)\text{Cos}[2\theta]+(-1+2a)(1+2\Lambda)^2\text{Cos}[2\theta]^2)}{\sqrt{16(-1+a)a\Lambda^2(1+\Lambda)^2+(1+2\Lambda(1+\Lambda)-(-1+2a)(1+2\Lambda)\text{Cos}[2\theta])^2}}$$
$$+\frac{2(4(-1+2a)\Lambda^2(1+\Lambda)^2+(1+4\Lambda+6\Lambda^2+4\Lambda^3)\text{Cos}[2\theta]+(-1+2a)(1+2\Lambda)^2\text{Cos}[2\theta]^2)}{\sqrt{16(-1+a)a\Lambda^2(1+\Lambda)^2+(1+2\Lambda(1+\Lambda)+(-1+2a)(1+2\Lambda)\text{Cos}[2\theta])^2}} = 0 \tag{19}$$

One can verify that the above reduces identically to zero when $a = 1/2$, for all $0 \leq \Lambda \leq \infty$, $0 \leq \theta \leq \frac{\pi}{4}$ and that furthermore $a = 1/2$ corresponds to a minimum of $\hat{H}(a, \Lambda, \theta)$ where $\frac{\partial^2 \hat{H}(a,\Lambda,\theta)}{\partial a^2} > 0$ can be shown. Furthermore, it is seen from Eq. (17) that this reduces to

$$\hat{H}(1/2, \Lambda, \theta) = 4\Lambda \tag{20}$$

Hence $\hat{H}(a, \Lambda, \theta) \geq \hat{H}(1/2, \Lambda, \theta) = 4\Lambda \geq 0$ and the result $\bar{E} + D \leq 1$ is proven for all pure initial states.

Note that when the initial state is a maximally entangled initial state, the inequality $\bar{E} + D \leq 1$ can be rewritten as $\bar{E} + D \leq E_i$, where $E_i$ is the initial entanglement of Qubits A and B. After Alice's measurement the change in entanglement from the initial to the final is given by $\Delta \bar{E} = \bar{E} - E_i$. Upon defining the average entanglement loss $\bar{E}_L \equiv -\Delta \bar{E}$ between the initial entanglement and final average entanglement, we have the direct relationship between distortion and entanglement loss $D \leq \bar{E}_L$. That is, for initial maximally entangled states as Alice causes distortion $D$ in order to obtain information about a particle in the process of measurement, such measurement inevitably leads to a corresponding loss of entanglement between Alice's particle and Bob's particle. An example application of this loss in the area of quantum eavesdropping will be examined in Sec. 4 of the conclusions.

*3.2. Two entangled particles, measurements by Alice and Bob*

In the prior section, two entangled particles of Alice and Bob were considered whereby Alice makes a weak measurement on her qubit. However, no measurement was done by Bob in the prior section. In this subsection, measurements by both Alice and Bob are considered. Alice first makes a measurement in the basis $\{M_0^{(A)}, M_1^{(A)}\}$. Bob's task is to attempt to determine as best as possible in the sense of minimizing the probability of error, what Alice's measurement result was. It is assumed that Bob knows the basis $\{M_0^{(A)}, M_1^{(A)}\}$ that Alice uses in her measurement, but not Alice's result.

We desire to determine the tradeoff between the average entanglement and the average information gain, and will show that $\bar{E} + \bar{G} \leq 1$. In order to determine $\bar{G}$, the minimum average probability of error will need to be determined. It was found that by substituting $b = \sin^2 \theta$ and $1 - b = \cos^2 \theta$ into Eq. (5), the resulting expressions for the two-particle measurement problem are simplified. Previously $0 \leq \theta \leq \frac{\pi}{4}$ which corresponds to $0 \leq b \leq 1/2$. Hence, in this section the following equivalent form for the measurement operators are utilized:

$$M_0^{(A)} \equiv \frac{1}{\sqrt{1 + 2\Lambda(1 + \Lambda)}} \begin{bmatrix} \Lambda + 1 - b & e^{i\phi}\sqrt{(1-b)b} \\ e^{-i\phi}\sqrt{(1-b)b} & \Lambda + b \end{bmatrix}$$

$$M_1^{(A)} \equiv \frac{1}{\sqrt{1 + 2\Lambda(1 + \Lambda)}} \begin{bmatrix} \Lambda + b & -e^{i\phi}\sqrt{(1-b)b} \\ -e^{-i\phi}\sqrt{(1-b)b} & \Lambda + 1 - b \end{bmatrix}.$$

(21)

As in the prior section, the initial state is given by $|\psi_0\rangle = \sqrt{\alpha_0}|0\rangle \otimes |0\rangle + \sqrt{1 - \alpha_0}|1\rangle \otimes |1\rangle$. Several equations for the case of two-particle measurement are the same as one-particle measurement. The equivalent forms for Eqs. (11)-(14) are found by substituting $1 - 2b$ for $\text{Cos}[2\theta]$ and $2\sqrt{b(1-b)}$ for $\text{Sin}[2\theta]$.

Now, after Alice makes her measurement, there are two possible outcomes. Bob desires to determine Alice's measurement as best as possible in the sense of minimizing the probability of error of his measurement outcome compared with Alice's outcome. After Alice's measurement, Bob will have one of two possible density matrices $\rho_{B,0}$ or $\rho_{B,1}$ as given in Eq. (13). The objective of Bob is to measure his density matrix in a manner that his outcome maximally correlates with Alice's so as to minimize the average probability of error. Bob, knowing the initial state and Alice's measurement operators but not Alice's outcome, needs to find such optimal measurement operators $\{M_0^{(B)}, M_1^{(B)}\}$. Bob's measurement operators that optimally discriminate between $\rho_{B,0}$ or $\rho_{B,1}$ were found theoretically by Bergou [29]. The optimal solution is found by first constructing the matrix $T$,

$$T \equiv \pi_{A,1}\rho_{B,1} - \pi_{A,0}\rho_{B,0} \tag{22}$$

and decomposing $T = \sum_{k=1}^{D_S} \lambda_{T,k}|\phi_k\rangle\langle\phi_k|$ where the states $\phi_k$ denote orthonormal eigenstates corresponding to the eigenvalues $\lambda_{T,k}$ of $T$. Bergou assumes the eigenvalues are numbered in the following manner:

$$\lambda_{T,k} < 0 \text{ for } 1 \leq k \leq k_0$$

$$\lambda_{T,k} > 0 \text{ for } k_0 \leq k \leq D$$

$$\lambda_{T,k} = 0 \text{ for } D \leq k \leq D_S. \tag{23}$$

The optimal POVMs are found by Bergou to be given by

$$\Pi_1 = \sum_{k=1}^{k_0-1} |\phi_k\rangle\langle\phi_k|$$

$$\Pi_2 = \sum_{k=k_0}^{D_S} |\phi_k\rangle\langle\phi_k|. \tag{24}$$

where $\Pi_1 + \Pi_2 = I$, and the expression for $\Pi_2$ has been supplemented by orthogonal eigenstates corresponding to the eigenvalues $\lambda_{T,k} = 0$.

Now, $T$ can be written in closed form via Eqs. (11)-(13) as:

$$T = \begin{pmatrix} \dfrac{a(-1+2b)(1+2\Lambda)}{1+2\Lambda(1+\Lambda)} & -\dfrac{2\sqrt{(-1+a)a(-1+b)b}e^{-i\phi}(1+2\Lambda)}{1+2\Lambda(1+\Lambda)} \\ -\dfrac{2\sqrt{(-1+a)a(-1+b)b}e^{i\phi}(1+2\Lambda)}{1+2\Lambda(1+\Lambda)} & \dfrac{(-1+a)(-1+2b)(1+2\Lambda)}{1+2\Lambda(1+\Lambda)} \end{pmatrix}. \tag{25}$$

The eigenstates are given by

$$|\phi_1\rangle = \begin{pmatrix} \sqrt[4]{\dfrac{(-1+a)a(-1+b)b}{1+\dfrac{16(-1+a)a(-1+b)b}{(-1+2b+\sqrt{1-4(1-2a)^2 b+4(1-2a)^2 b^2})^2}}} e^{-i\phi} \\ \dfrac{1}{\sqrt{1+\dfrac{16(-1+a)a(-1+b)b}{(-1+2b+\sqrt{1-4(1-2a)^2 b+4(1-2a)^2 b^2})^2}}} \end{pmatrix},$$

$$|\phi_2\rangle = \begin{pmatrix} -\dfrac{\sqrt[4]{\dfrac{(-1+a)a(-1+b)b}{1+\dfrac{16(-1+a)a(-1+b)b}{(1-2b+\sqrt{1-4(1-2a)^2 b+4(1-2a)^2 b^2})^2}}} e^{-i\phi}}{1-2b+\sqrt{1-4(1-2a)^2 b+4(1-2a)^2 b^2}} \\ \dfrac{1}{\sqrt{1+\dfrac{16(-1+a)a(-1+b)b}{(1-2b+\sqrt{1-4(1-2a)^2 b+4(1-2a)^2 b^2})^2}}} \end{pmatrix}$$

(26)

with corresponding eigenvalues

$$\lambda_{T,1} = \frac{(1-2a-2b+4ab-\sqrt{-4(-a+a^2)+(-1+2a+2b-4ab)^2})(1+2\Lambda)}{2(1+2\Lambda+2\Lambda^2)}$$

$$\lambda_{T,2} = \frac{(1-2a-2b+4ab+\sqrt{-4(-a+a^2)+(-1+2a+2b-4ab)^2})(1+2\Lambda)}{2(1+2\Lambda+2\Lambda^2)}$$

(27)

One can see by inspection that $\lambda_{T,1} \leq 0$ and $\lambda_{T,2} \geq 0$ with $0 \leq a \leq \frac{1}{2}, 0 \leq b \leq \frac{1}{2}$, hence the conditions established in Bergou are met and the optimal POVM elements $\Pi_0, \Pi_1$ can be computed as

$$\Pi_0 = \begin{pmatrix} \dfrac{1-2b+\sqrt{1+4(1-2a)^2(-1+b)b}}{2\sqrt{1+4(1-2a)^2(-1+b)b}} & \dfrac{\sqrt[4]{\dfrac{(-1+a)a(-1+b)b}{(1+\dfrac{16(-1+a)a(-1+b)b}{(-1+2b+\sqrt{1+4(1-2a)^2(-1+b)b})^2})^2}} e^{-i\phi}}{-1+2b+\sqrt{1+4(1-2a)^2(-1+b)b}} \\ \dfrac{\sqrt[4]{\dfrac{(-1+a)a(-1+b)b}{(1+\dfrac{16(-1+a)a(-1+b)b}{(-1+2b+\sqrt{1+4(1-2a)^2(-1+b)b})^2})^2}} e^{i\phi}}{-1+2b+\sqrt{1+4(1-2a)^2(-1+b)b}} & \dfrac{1}{1+\dfrac{16(-1+a)a(-1+b)b}{(1-2b+\sqrt{1+4(1-2a)^2(-1+b)b})^2}} \end{pmatrix}$$

(28)

$$\Pi_1 = \begin{pmatrix} \dfrac{-1+2b+\sqrt{1+4(1-2a)^2(-1+b)b}}{2\sqrt{1+4(1-2a)^2(-1+b)b}} & -\dfrac{\sqrt[4]{(1+\dfrac{16(-1+a)a(-1+b)b}{(1-2b+\sqrt{1+4(1-2a)^2(-1+b)b})^2})^2}}{1-2b+\sqrt{1+4(1-2a)^2(-1+b)b}} e^{-i\phi} \\ -\dfrac{\sqrt[4]{(1+\dfrac{16(-1+a)a(-1+b)b}{(1-2b+\sqrt{1+4(1-2a)^2(-1+b)b})^2})^2}}{1-2b+\sqrt{1+4(1-2a)^2(-1+b)b}} e^{i\phi} & 1+\dfrac{1}{\dfrac{16(-1+a)a(-1+b)b}{(1-2b+\sqrt{1+4(1-2a)^2(-1+b)b})^2}} \end{pmatrix}.$$

From these equations, $\{M_0^{(B)}, M_1^{(B)}\}$ can be computed by decomposing $\Pi_i = M_i^{(B)\dagger} M_i^{(B)}$ (there may be more than a single implementation of $\{M_0^{(B)}, M_1^{(B)}\}$ that gives the optimal POVM $\{\Pi_0, \Pi_1\}$). Now, recall that the gain is $G_i \equiv 1 - 2P_{e|A=i}$, where $P_{e|A=i}$ is the probability of error given Alice obtains result $i \in \{0,1\}$ and the average gain is found as $\bar{G} = \pi_{A,0} G_0 + \pi_{A,1} G_1$. We can then compute $P_{e|A=0} = \text{Tr}[\Pi_1 \rho_{B,1}]$ and $P_{e|A=1} = \text{Tr}[\Pi_0 \rho_{B,2}]$. Upon substituting the closed form expressions above, we obtain:

$$\begin{aligned}
P_{e|A=0} &= \frac{1}{2}\left(1 + \frac{-8a^2(-1+b)b(1+2\Lambda) - (-1+2b)(b+2b\Lambda+\Lambda^2) + a(-1-6b+4b(-3+\Lambda)\Lambda - 2\Lambda(1+\Lambda) + 8b^2(1+2\Lambda))}{\sqrt{1+4(1-2a)^2(-1+b)b}(b+2b\Lambda+\Lambda^2 - a(-1+2b)(1+2\Lambda))}\right) \\
P_{e|A=1} &= \frac{1}{2}\left(1 - \frac{8a^2(-1+b)b(1+2\Lambda) + (-1+2b)(b+2b\Lambda-(1+\Lambda)^2) - a(1+2\Lambda+2\Lambda^2 + 8b^2(1+2\Lambda) - 2b(5+10\Lambda+2\Lambda^2))}{\sqrt{1+4(1-2a)^2(-1+b)b}((1+\Lambda)^2 - b(1+2\Lambda) + a(-1+2b)(1+2\Lambda))}\right).
\end{aligned} \quad (29)$$

The information gains are now computed as

$$\begin{aligned}
G_0 &= -\frac{-8a^2(-1+b)b(1+2\Lambda) - (-1+2b)(b+2b\Lambda+\Lambda^2) + a(-1-6b+4b(-3+\Lambda)\Lambda - 2\Lambda(1+\Lambda) + 8b^2(1+2\Lambda))}{\sqrt{1+4(1-2a)^2(-1+b)b}(b+2b\Lambda+\Lambda^2 - a(-1+2b)(1+2\Lambda))} \\
G_1 &= \frac{8a^2(-1+b)b(1+2\Lambda) + (-1+2b)(b+2b\Lambda-(1+\Lambda)^2) - a(1+2\Lambda+2\Lambda^2 + 8b^2(1+2\Lambda) - 2b(5+10\Lambda+2\Lambda^2))}{\sqrt{1+4(1-2a)^2(-1+b)b}((1+\Lambda)^2 - b(1+2\Lambda) + a(-1+2b)(1+2\Lambda))}
\end{aligned} \quad (30)$$

Note that $\bar{E} + \bar{G} = \pi_{A,0} E(\rho_{B,0}) + \pi_{A,1} E(\rho_{B,1}) + \pi_{A,0} G_0 + \pi_{A,1} G_1 = \pi_{A,0}(E(\rho_{B,0}) + G_0) + \pi_{A,1}(E(\rho_{B,1}) + G_1)$. Using $E(\rho_{B,i})$ from (14) and $G_i$ from above with $i = 0,1$, the following is obtained:

$$\begin{aligned}
&E(\rho_{B,0}) + G_0 \\
&= -\frac{-8a^2(-1+b)b(1+2\Lambda) - (-1+2b)(b+2b\Lambda+\Lambda^2) + a(-1-6b+4b(-3+\Lambda)\Lambda - 2\Lambda(1+\Lambda) + 8b^2(1+2\Lambda))}{\sqrt{1+4(1-2a)^2(-1+b)b}(b+2b\Lambda+\Lambda^2 - a(-1+2b)(1+2\Lambda))} \\
&\quad + 2\left(\frac{1}{2} + \frac{\sqrt{16(-1+a)a\Lambda^2(1+\Lambda)^2 + (1+2\Lambda(1+\Lambda) + (-1+2a)(1-2b)(1+2\Lambda))^2}}{4a(-1+2b)(1+2\Lambda) - 4(b+2b\Lambda+\Lambda^2)}\right)
\end{aligned} \quad (31)$$

$$
\begin{aligned}
&E(\rho_{B,1}) + G_1 \\
&= \frac{8a^2(-1+b)b(1+2\Lambda) + (-1+2b)(b+2b\Lambda - (1+\Lambda)^2) - a(1+2\Lambda(1+\Lambda) + 8b^2(1+2\Lambda) - 2b(5+2\Lambda(5+\Lambda)))}{\sqrt{1+4(1-2a)^2(-1+b)b}((1+\Lambda)^2 - b(1+2\Lambda) + a(-1+2b)(1+2\Lambda))} \\
&\quad + 2\left(\frac{1}{2} - \frac{\sqrt{16(-1+a)a\Lambda^2(1+\Lambda)^2 + \left(1+2\Lambda(1+\Lambda) + (-1+2a)(-1+2b)(1+2\Lambda)\right)^2}}{2+4\Lambda(1+\Lambda) + 2(-1+2a)(-1+2b)(1+2\Lambda)}\right).
\end{aligned}
$$

Upon substituting for $\pi_{A,i}$ from Eq. (11) ($i = 0,1$) into $\pi_{A,0}\big(E(\rho_{B,0}) + G_0\big) + \pi_{A,1}\big(E(\rho_{B,1}) + G_1\big)$ and reducing, then $\pi_{A,0}\big(E(\rho_{B,0}) + G_0\big) + \pi_{A,1}\big(E(\rho_{B,1}) + G_1\big) = \bar{E} + \bar{G} - 1$ is obtained as:

$$
\frac{f_1(a,b,\Lambda) - f_2(a,b,\Lambda) - f_3(a,b,\Lambda)}{\sqrt{1+4(1-2a)^2(-1+b)b}\,(1+2\Lambda(1+\Lambda))} \tag{32}
$$

where

$$
f_1(a,b,\Lambda) = (1 + 2\Lambda + 4(1-2a)^2(-1+b)b(1+2\Lambda))
$$

$$
\begin{aligned}
f_2(a,b,\Lambda) = &[(1+4(1-2a)^2(-1+b)b)((a+b-2ab)^2 + 4(a+b-2ab)^2\Lambda + 2(a(-1+4a) + \\
&b + 2(1-4a)ab + 2(1-2a)^2 b^2)\Lambda^2 + 4(-1+2a)(a-b)\Lambda^3 + (1-2a)^2\Lambda^4)]^{1/2}
\end{aligned}
$$

$$
\begin{aligned}
f_3(a,b,\Lambda) = &[(1+4(1-2a)^2(-1+b)b)((b+2b\Lambda - (1+\Lambda)^2)^2 + a^2(-4b(1+2\Lambda)^2 + 4(b+2b\Lambda)^2 + \\
&(1+2\Lambda(1+\Lambda))^2) - 2a(2(b+2b\Lambda)^2 + (1+\Lambda)^2(1+2\Lambda(1+\Lambda)) - b(1+2\Lambda)(3+2\Lambda(3+\Lambda))]^{\frac{1}{2}}.
\end{aligned}
$$

Now, we desire to prove that $\bar{E} + \bar{G} \leq 1$ or $\bar{E} + \bar{G} - 1 \leq 0$. Note that the prefactor $\left(\sqrt{1+4(1-2a)^2(-1+b)b}(1+2\Lambda(1+\Lambda))\right)^{-1}$ is always greater than or equal to zero when $0 \leq a \leq \frac{1}{2}, 0 \leq b \leq \frac{1}{2}$. Hence, we can multiply both sides of $\bar{E} + \bar{G} - 1 \leq 0$ by the prefactor obtaining the inequality

$$
f_1(a,b,\Lambda) - f_2(a,b,\Lambda) - f_3(a,b,\Lambda) \leq 0 \tag{33}
$$

We desire to determine if $f_1 - f_2 - f_3 \leq 0$. In order to determine this, consider the following lemma that is easily proven. Let $h(x)$ and $g(x)$ be real functions defined on a set $x \in S \subset \mathbb{R}$, where $\mathbb{R}$ denotes the set of real numbers. Suppose $h(x) \geq 0, x \in S$. If $h^2(x) \geq g^2(x)$ then $h(x) \geq g(x)$. Now, the above inequality can be rewritten $f_2 + f_3 \geq f_1$. It can be seen that $f_2 \geq 0, f_3 \geq 0, f_2 + f_3 \geq 0$ hence if $(f_2 + f_3)^2 \geq f_1^2$ then it follows that $f_2 + f_3 \geq f_1$.

Now, $(f_2 + f_3)^2 - f_1^2 \geq 0$ can be written as $-(f_2 + f_3)^2 + f_1^2 \leq 0$ or expanding this expression:

$$
\begin{aligned}
&-2(1+4(1-2a)^2(-1+b)b)\Big[\Lambda^2(1+\Lambda)^2 + b(1+2\Lambda)^2 - (b+2b\Lambda)^2 + a\Big(-4b(1+2\Lambda)^2 + 4(b+2b\Lambda)^2 - \\
&(1+2\Lambda(1+\Lambda))^2\Big) + a^2\Big(4b(1+2\Lambda)^2 - 4(b+2b\Lambda)^2 + \big(1+2\Lambda(1+\Lambda)\big)^2\Big) + (((a+b-2ab)^2 + 4(a+b-2ab)^2\Lambda + \\
&2(a(-1+4a) + b + 2(1-4a)ab + 2(1-2a)^2 b^2)\Lambda^2 + 4(-1+2a)(a-b)\Lambda^3 + (1-2a)^2\Lambda^4)((-1+a+b-2ab)^2 + \\
&4(-1+a+b-2ab)^2\Lambda + 2(3+a(-7+4a) - 5b + 2(7-4a)ab + 2(1-2a)^2 b^2)\Lambda^2 + 4(-1+2a)(-1+a+b)\Lambda^3 + \\
&(1-2a)^2\Lambda^4))^{1/2}\Big] \leq 0
\end{aligned}
$$

The prefactor $-2(1 + 4(1 - 2a)^2(-1 + b)b)$ is always less than or equal to zero when $0 \leq a \leq \frac{1}{2}, 0 \leq b \leq \frac{1}{2}$. Dividing both sides by the prefactor the following inequality is obtained:

$$-2(1 + 4(1 - 2a)^2(-1 + b)b)\Big[\Lambda^2(1 + \Lambda)^2 + b(1 + 2\Lambda)^2 - (b + 2b\Lambda)^2 + a\Big(-4b(1 + 2\Lambda)^2 + 4(b + 2b\Lambda)^2 - (1 + 2\Lambda(1 + \Lambda))^2\Big) + a^2\Big(4b(1 + 2\Lambda)^2 - 4(b + 2b\Lambda)^2 + (1 + 2\Lambda(1 + \Lambda))^2\Big) + (((a + b - 2ab)^2 + 4(a + b - 2ab)^2\Lambda + 2(a(-1 + 4a) + b + 2(1 - 4a)ab + 2(1 - 2a)^2b^2)\Lambda^2 + 4(-1 + 2a)(a - b)\Lambda^3 + (1 - 2a)^2\Lambda^4)((-1 + a + b - 2ab)^2 + 4(-1 + a + b - 2ab)^2\Lambda + 2(3 + a(-7 + 4a) - 5b + 2(7 - 4a)ab + 2(1 - 2a)^2b^2)\Lambda^2 + 4(-1 + 2a)(-1 + a + b)\Lambda^3 + (1 - 2a)^2\Lambda^4))^{1/2}\Big] \leq 0 \quad (34)$$

Denote the terms $h_1$ and $h_2$ as:

$$h_1 = \Lambda^2(1 + \Lambda)^2 + b(1 + 2\Lambda)^2 - (b + 2b\Lambda)^2 + a\Big(-4b(1 + 2\Lambda)^2 + 4(b + 2b\Lambda)^2 - (1 + 2\Lambda(1 + \Lambda))^2\Big) + a^2\Big(4b(1 + 2\Lambda)^2 - 4(b + 2b\Lambda)^2 + (1 + 2\Lambda(1 + \Lambda))^2\Big)$$

$$h_2 = (((a + b - 2ab)^2 + 4(a + b - 2ab)^2\Lambda + 2(a(-1 + 4a) + b + 2(1 - 4a)ab + 2(1 - 2a)^2b^2)\Lambda^2 + 4(-1 + 2a)(a - b)\Lambda^3 + v(1 - 2a)^2\Lambda^4)((-1 + a + b - 2ab)^2 + 4(-1 + a + b - 2ab)^2\Lambda + 2(3 + a(-7 + 4a) - 5b + 2(7 - 4a)ab + 2(1 - 2a)^2b^2)\Lambda^2 + 4(-1 + 2a)(-1 + a + b)\Lambda^3 + (1 - 2a)^2\Lambda^4))^{1/2}. \quad (35)$$

It is then desired to show that $h_1 + h_2 \geq 0$. Note that $h_2$ can be seen to be greater than or equal to zero when $0 \leq a \leq \frac{1}{2}, 0 \leq b \leq \frac{1}{2}$. Now, if $h_1^2 \geq (-h_2)^2$ then $h_1 \geq -h_2$. It can be verified that $h_1^2 + (-h_2)^2$ factorizes as follows:

$$h_1^2 + (-h_2)^2 = 4(1 - 2a)^2(-1 + a)a(-1 + b)b\Big(1 + 2\Lambda\big(2 + \Lambda(3 + 2\Lambda)\big)\Big)^2. \quad (36)$$

When $0 \leq a \leq \frac{1}{2}, 0 \leq b \leq \frac{1}{2}$ all terms in the RHS of the equation above are nonnegative. Hence $h_1^2 + (-h_2)^2 \geq 0$ and the proof that $\bar{E} + \bar{G} \leq 1$ is complete.

## 4. Conclusions

Assuming an initial entangled two-particle state shared by Alice and Bob, both measurement on a single system by Alice and measurement on both systems by both Alice and Bob have been considered in regards to the effect on the entanglement. For single-particle measurement by Alice, it has been proven that the sum of the entanglement and the measurement disturbance, quantified by $D$, is less than or equal to unity, $\bar{E} + D \leq 1$. For the case of two-particle measurement, it has been shown that both entanglement and the average information gained through measurement $\bar{G}$ cannot both be arbitrarily high, i.e. $\bar{E} + \bar{G} \leq 1$.

The results were initially proven over arbitrary initial pure states and extended to arbitrary mixed states in Appendix 2 of the Supplementary Material. As well, non-Hermitian measurement implementations of a given positive operator valued measurement were considered in Appendix 3 of the Supplementary Material and the complementary results proven to continue to hold.

In the case of maximally entangled states initial states, it is found that entanglement loss is directly complementary to both measurement disturbance and information gain. This provides a simple and easy to apply formula for such initial states that are widely utilized in both theory and experiment in quantum information which often utilize maximally entangled states.

One might consider what applications do such bounds have. Let us consider an example of the potential application of the bounds in the area of secure communication. Consider Alice desires to transmit classical information of a classically encrypted bit stream to Bob through a medium. Suppose Alice uses quantum degrees of freedom such as $|H\rangle$ and $|V\rangle$ photon polarization to represent classical bits of 0 and 1 respectively. Suppose that the medium is not secure and is subject being tapped by an eavesdropper, Eve. Eve is allowed to make measurements on the qubits and as well retransmit the qubits.

In the case when Alice transmits classical bits to Bob, Eve can intercept the bits, learn them, and then retransmit the bits without detection. It does not appear that at least classically, the presence or absence of an eavesdropper such as Eve can be detected by Alice and Bob since Eve retransmits the bits precisely as she receives them. Consider the following protocol which utilizes a simplified version of a quantum protocol proposed by Humble [30] to illustrate how Eve can be detected by Alice and Bob by use of quantum entanglement. Let us suppose that Alice will generate two polarization maximally entangled qubits at some rate, and interleaves one of these within her encrypted classical data stream in a manner at least initially only known to Alice and Bob. Alice and Bob then perform a Bell experiment on a given number of pairs to characterize the violation. Now, if Eve is eavesdropping, she will have measured some of these interleaved particles sent to Bob and we can apply our bound $D \leq \bar{E}_L$, where $\bar{E}_L \equiv -\Delta \bar{E}$ is the entanglement loss, or $\bar{E}_f \leq 1 - D$. If Eve makes strong measurements corresponding to $\Lambda \to 0$, or $D \to 1$ we see that $\bar{E}_f \to 0$ results immediately and the initial entanglement will be necessarily destroyed by Eve in the process of eavesdropping. Hence, Alice and Bob can reliably detect the presence of Eve if she uses strong measurement. Note that Eve can utilize weak measurement so as to minimize her disturbance. Given some background channel noise level it may be advantageous for Eve to lower her disturbance in a manner that makes it more difficult for Alice and Bob to detect her. This however will lower her ability to make reliable measurements of the channel. Other papers have considered this and as well other schemes including using continuous entanglement [31]. An experimental demonstration of confidential communication, albeit with a modified version of these techniques, was recently reported in [32]

## Acknowledgments

Mathematica was utilized in the derivation of the equations and Maple was used to check these algebraic results. MATLAB was utilized to double-check the closed-form equations numerically.

**Author Contributions:** Conceptualization, Michael Steiner and Ronald Rendell.; Investigation, Michael Steiner and Ronald Rendell.; writing— Michael Steiner and Ronald Rendell; writing—review and editing, Michael Steiner and Ronald Rendell. All authors have read and agreed to the published version of the manuscript.

**Funding:** This research received no external funding.

**Conflict of Interest**: The authors declare no conflict of interest.

**Data availability statement:** Not applicable

**Institutional review board statement:** Not applicable.

**Informed consent statement:** Not applicable

**Sample Statement**: The authors declare that no physical samples were used in this study.

**Supplementary materials:** The supplementary materials Appendices 1,2,3 are available at [Link to resource location]

# Supplement 1: Appendices 1, 2, 3, 4

## Appendix 1. Class of Initial Pure States

We first consider two-particle measurement. Consider an arbitrary initial pure state of the form $\sqrt{\alpha_0}|u_1\rangle \otimes |v_1\rangle + \sqrt{1-\alpha_0}|u_2\rangle \otimes |v_2\rangle$ with $|u_1\rangle \perp |u_2\rangle, |v_1\rangle \perp |v_2\rangle$, for which it is desired to prove that $\bar{E} + \bar{G} \leq 1$ given Alice applies arbitrary positive-semidefinite operators $\widehat{M}_i^{(A)} \in S^2, i = 0,1$ where $S^2$ denotes the set of $2 \times 2$ positive semi-definite matrices of the form of

$$M_0^{(A)} \equiv \frac{1}{\sqrt{1 + 2\Lambda(1 + \Lambda)}} \begin{bmatrix} \Lambda + \cos^2 \theta & e^{i\phi} \sin \theta \cos \theta \\ e^{-i\phi} \sin \theta \cos \theta & \Lambda + \sin^2 \theta \end{bmatrix}$$

$$M_1^{(A)} \equiv \frac{1}{\sqrt{1 + 2\Lambda(1 + \Lambda)}} \begin{bmatrix} \Lambda + \sin^2 \theta & -e^{i\phi} \sin \theta \cos \theta \\ -e^{-i\phi} \sin \theta \cos \theta & \Lambda + \cos^2 \theta \end{bmatrix} \quad (1.1)$$

are applied by Alice. It is assumed, for the class of problems for which any $M_i^{(A)} \in S^2, i = 0,1$, is applied by Alice to the initial state $\sqrt{\alpha_0}|0\rangle \otimes |0\rangle + \sqrt{1-\alpha_0}|1\rangle \otimes |1\rangle$, that there exists an optimal measurement $\{M_0^{(B)}, M_1^{(B)}\}$ applied by Bob that results in $\bar{E} + \bar{G} \leq 1$.

Consider now the following lemma. Suppose that for any initial pure state, Alice applies operators $\{M_0^{(A)}, M_1^{(A)}\}$ to her qubit for which $M_0^{(A)\dagger} M_0^{(A)} + M_1^{(A)\dagger} M_1^{(A)} = I$, and then Bob applies optimal operators $\{M_0^{(B)}, M_1^{(B)}\}$ to his qubit and finds $s_1 = \bar{E} + \bar{G} \leq 1$. Then for any unitary operator $U_a$, 1) the set $\{U_a M_0^{(A)}, U_a M_1^{(A)}\}$ also constitutes a valid POVM implementation, and 2) if Alice instead applies operators $\{U_a M_0^{(A)}, U_a M_1^{(A)}\}$ to her qubit, then $\bar{E} + \bar{G} = s_1 \leq 1$ is invariant. This is because $\pi_{A,0}, \rho_{B,0}, \pi_{A,1}, \rho_{B,1}$ are all independent of $U_a$ while both $\bar{E}$ and $\bar{G}$ are functions of only these quantities.

Consider again an arbitrary initial pure state of the form $\sqrt{\alpha_0}|u_1\rangle \otimes |v_1\rangle + \sqrt{1-\alpha_0}|u_2\rangle \otimes |v_2\rangle$ for which it is desired to prove that $\bar{E} + \bar{G} \leq 1$ given Alice applies arbitrary positive-semidefinite operators $\widehat{M}_i^{(A)} \in S^2, i = 0,1$. Define the local unitary $U_B \otimes U_A$ for which $U_B$ maps $|u_1\rangle \to |0\rangle, |u_2\rangle \to |1\rangle$ and $U_A$ maps $|v_1\rangle \to |0\rangle, |v_2\rangle \to |1\rangle$. Note that when the operators $\{\widehat{M}_0^{(A)} U_A, \widehat{M}_1^{(A)} U_A\}$ are applied by Alice to the initial state $\sqrt{\alpha_0}|0\rangle \otimes |0\rangle + \sqrt{1-\alpha_0}|1\rangle \otimes |1\rangle$, this is identical to Alice applying $\{\widehat{M}_0^{(A)}, \widehat{M}_1^{(A)}\}$ to the initial state $\sqrt{\alpha_0}|u_1\rangle \otimes |v_1\rangle + \sqrt{1-\alpha_0}|u_2\rangle \otimes |v_2\rangle$. However, the operators $\{\widehat{M}_0^{(A)} U_A, \widehat{M}_1^{(A)} U_A\}$ are not necessarily in $S^2$. If Alice instead applies $\{U_A^\dagger \widehat{M}_0^{(A)} U_A, U_A^\dagger \widehat{M}_1^{(A)} U_A\}$, such matrices are in $S^2$. From the lemma above, it is known that $\bar{E} + \bar{G}$ is unchanged with the addition of the unitary $U_A^\dagger$ to the LHS of the operators. Defining $M_0^{(A)} \equiv U_A^\dagger \widehat{M}_0^{(A)} U_A$ and $M_1^{(A)} \equiv U_A^\dagger \widehat{M}_1^{(A)} U_A$ and noting that $M_i^{(A)} \in S^2, i = 0,1$ and the initial state is $\sqrt{\alpha_0}|0\rangle \otimes |0\rangle + \sqrt{1-\alpha_0}|1\rangle \otimes |1\rangle$, this falls within the class of problems for which $\bar{E} + \bar{G} \leq 1$. This shows that once it is proven that with initial states $\sqrt{\alpha_0}|0\rangle \otimes |0\rangle + \sqrt{1-\alpha_0}|1\rangle \otimes |1\rangle$ that $\bar{E} + \bar{G} \leq 1$, it implies that any initial pure state can be utilized and $\bar{E} + \bar{G} \leq 1$. Hence one can consider for pure states, without loss of generality, pure states of the form $\sqrt{\alpha_0}|0\rangle \otimes |0\rangle + \sqrt{1-\alpha_0}|1\rangle \otimes |1\rangle$ in the proof of $\bar{E} + \bar{G} \leq 1$.

Note that aspects of the argument above can also be applied to single-particle measurement. Consider again an arbitrary initial pure state of the Schmidt form $\sqrt{\alpha_0}|u_1\rangle \otimes |v_1\rangle + \sqrt{1-\alpha_0}|u_2\rangle \otimes |v_2\rangle$ with $|u_1\rangle \perp |u_2\rangle, |v_1\rangle \perp |v_2\rangle$ for which it is desired to prove that $\bar{E} + D \leq 1$ given Alice applies arbitrary positive-semidefinite operators $\widehat{M}_i^{(A)} \in S^2, i = 0,1$. If we apply the local unitary $U_B \otimes U_A$ for which $U_B$ maps $|u_1\rangle \to |0\rangle, |u_2\rangle \to |1\rangle$ and $U_A$ maps $|v_1\rangle \to |0\rangle, |v_2\rangle \to |1\rangle$ then the state becomes $|\psi_0\rangle = \sqrt{a}|0\rangle \otimes |0\rangle + \sqrt{1-a}|1\rangle \otimes |1\rangle$. Similar to the above, when the operators $\{\widehat{M}_0^{(A)}U_A, \widehat{M}_1^{(A)}U_A\}$ are applied by Alice to the initial state $\sqrt{\alpha_0}|0\rangle \otimes |0\rangle + \sqrt{1-\alpha_0}|1\rangle \otimes |1\rangle$, this is identical to Alice applying $\{\widehat{M}_0^{(A)}, \widehat{M}_1^{(A)}\}$ to the initial state $\sqrt{\alpha_0}|u_1\rangle \otimes |v_1\rangle + \sqrt{1-\alpha_0}|u_2\rangle \otimes |v_2\rangle$. However, the operators $\{\widehat{M}_0^{(A)}U_A, \widehat{M}_1^{(A)}U_A\}$ are not necessarily in $S^2$. If Alice instead applies $\{U_A^\dagger \widehat{M}_0^{(A)}U_A, U_A^\dagger \widehat{M}_1^{(A)}U_A\}$, such matrices are in $S^2$. Note that as in the previous two-particle proof, $\bar{E}$ is left unchanged. Hence without loss of generality for the same reasons as above, we consider initial states of the form $|\psi_0\rangle = \sqrt{a}|0\rangle \otimes |0\rangle + \sqrt{1-a}|1\rangle \otimes |1\rangle$.

## Appendix 2. Initial Mixed States

We will look at extending both the single-particle measurement theory of Sec. 3.1 and the two-particle measurement theory of Sec. 3.2 from an initial pure state to a mixed state. In either case, we consider the use of the entanglement of formation as entanglement quantification for mixed states. Given any decomposition of $\rho = \sum_{i=1}^N \lambda_k |\phi_k\rangle\langle\phi_k|$, the entanglement of formation [1] $E_F \equiv \min_{|\phi_k\rangle} \sum_{i=1}^N \lambda_k E_H(|\phi_k\rangle\langle\phi_k|)$ where $E_H(|\phi_k\rangle\langle\phi_k|)$ denotes the entropy entanglement of the pure state $|\phi_k\rangle$. In this section, we will denote a generalized entanglement of formation relative to any entanglement measure $E$ (defined on pure states) as $E_{FG} \equiv \min_{|\phi_k\rangle} \sum_{i=1}^N \lambda_k E(|\phi_k\rangle\langle\phi_k|)$ and the states $|\phi_k\rangle$ that achieve the minimum in the entanglement of formation are denoted $|\phi_k^o\rangle$ and the respective coefficients in the expansion of entanglement of formation denoted $\lambda_k^o$. In this paper, the entanglement measure $E$ has been taken for bipartite pure-state systems as defined in Sec. 2.2, i.e. twice the minimum eigenvalue of the reduced density matrix. We will continue using this measure for pure states, and the entanglement of formation $E_{FG}$ relative to the latter pure state measure will be used for mixed states.

### 2.1. Single-particle Measurement

In the case of pure state single-particle measurement, Bob and Alice share an entangled pure state for which Alice's state unitarily interacts with a pointer detector qubit. The overlap $F = \langle \psi_{P,\downarrow} | \psi_{P,\uparrow} \rangle$ between the two pointer readings corresponding to spin-up and spin-down is related to the measurement strength. If the overlap $F$ is large, then little is learned of the actual spin by the weak measurement. If $F$ is small, then the strength of the measurement is high. We will now consider extending the prior results in Sec 2.4.1 to an initial system mixed state of Bob and Alice $\rho_{BA}$, and we will take our initial system-detector state to be $\rho_S \otimes |0\rangle\langle 0|$ where the system qubit is Alice's qubit, i.e. $\rho_S = \text{Tr}_B \rho_{AB}$.

Now, the results of the measurement disturbance in Sec. 3.1 are equally applicable to a mixed initial system state and $D = 1 - F = \frac{1}{1+2\Lambda(1+\Lambda)}$. Hence we need to address the effect an initial mixed state has on the entanglement as a function of the measurement parameters. Toward this end, suppose that unbeknownst to Alice and Bob, Charles can either initialize the joint Bob-Alice density matrix $\rho_{BA}$ qubit state to an initial improper mixed system state $\sum_{i=1}^N \lambda_k |\phi_k\rangle\langle\phi_k|$, or Bob-Alice's pure initial state $|\psi_0\rangle = \sqrt{\alpha_0}|u_1\rangle \otimes |v_1\rangle + \sqrt{1-\alpha_0}|u_2\rangle \otimes |v_2\rangle$ is initialized by Charles to one of the $|\phi_k\rangle$. In the

latter case, Charles randomly chooses the $k$ th vector $|\phi_k\rangle$ with probability $\pi_{C,k} = \lambda_k$. We will refer to this case as Charles generating a pure state ensemble $|\phi_k\rangle$ of $\rho_{BA}$.

We consider first the case where Charles initializes the composite state of Bob-Alice to a pure state and chooses the state $|\psi_0\rangle = |\phi_k\rangle$ with probability $\pi_{C,k} = \lambda_k$. Alice's qubit can then be considered to unitarily interact with a detector qubit that is then projected in the computational basis, as shown in Sec. 2.4.1. This has been noted to be identically equivalent to Alice applying weak measurement operators represented by $E_i = M_i^{(A)\dagger} M_i^{(A)}$, $i = 0,1$ with the $M_i^{(A)}$ as defined in Eq. (1.1), to her qubit that forms a POVM with $\sum E_i = I$. Hence we will consider the interaction of Alice's qubit with the detector and subsequent projection in the computational basis to be replaced by Alice simply applying to her qubit the weak measurement operators $E_i = M_i^{(A)\dagger} M_i^{(A)}$, $i = 0,1$.

The probabilities of Alice's outcomes are given by

$$\pi_{A,i} = \text{Tr}\left[|\psi_0\rangle\langle\psi_0|(I\otimes M_i^{(A)})^\dagger(I\otimes M_i^{(A)})\right] \tag{2.1}$$

and let us denote the resulting Bob-Alice state as $|\psi\rangle_{BA|i}$ and respective density matrix as $\rho_{BA|i}$ given Alice's outcome is $i \in \{0,1\}$. Furthermore, given Alice's outcome is $i \in \{0,1\}$ and given Charles's state selection is $|\psi_0\rangle = |\phi_k\rangle$, we denote the pure composite Bob-Alice state as $|\psi\rangle_{BA|i,k}$ with respective density matrix $\rho_{BA|i,k}$. The entanglement averaged over Alice's potential outcomes $i \in \{0,1\}$ is given by

$$\bar{E}(|\phi_k\rangle) = p(A=0|c=k)E(\rho_{BA|0,k}) + p(A=1|c=k)E(\rho_{BA|1,k}) \tag{2.2}$$

where $p(A=i|c=k)$ denotes the probability that Alice measures outcome $i \in \{0,1\}$ given that Charles chooses state $k \in \{1, \ldots, N\}$. For simplicity we will take $N = 2$, but the results are independent of $N$ so that the maximum rank of two qubits, $N = 4$ can be taken if desired. We already know from Sec. 3.1 that

$$\bar{E}(|\phi_k\rangle) + D \leq 1, k \in \{1, \ldots, N\}. \tag{2.3}$$

Now, let us consider the case that Charles initializes the system to the improper mixed state $\rho_{BA} = \sum_{i=1}^N \lambda_k |\phi_k\rangle\langle\phi_k|$. In this case, we denote the initial mixed density operator of Bob-Alice as $\rho_{BA}^{(x)}$ where $x$ denotes that Charles initially does not choose a pure state $|\phi_k\rangle$ by using a random number generator, but rather inserts the mixed improper state $\rho_{BA}^{(x)} = \sum_{i=1}^N \lambda_k |\phi_k\rangle\langle\phi_k|$. We desire to prove that $\bar{E}_F\left(\rho_{BA}^{(x)}\right) + D \leq 1$ where the entanglement of formation averaged over Alice's outcomes is denoted $\bar{E}_F\left(\rho_{BA}^{(x)}\right)$, and $D$ is the measurement disturbance. The average entanglement of formation is given by

$$\bar{E}_F\left(\rho_{BA}^{(x)}\right) = \pi_{A,0} E_F\left(\rho_{BA|0}^{(x)}\right) + \pi_{A,1} E_F(\rho_{BA|1}^{(x)}) \tag{2.4}$$

$$= \pi_{A,0}\big(E_F(\,p(c=1|A=0)\rho_{BA|0,1} + p(c=2|A=0)\rho_{BA|0,2}\big) \\ + \pi_{A,1}\big(E_F(\,p(c=1|A=1)\rho_{BA|1,1} + p(c=2|A=1)\rho_{BA|1,2}\big). \tag{2.5}$$

Now, in the equation above, $p(c = j|A = i)$, $k \in \{1,2\}$, $i \in \{0,1\}$, denotes the *a posteriori* probability that Charles had chosen pure state $|\phi_k\rangle$ given that Alice obtained outcome $i$ and $M_i^{(A)}$ was applied to her qubit. Although not necessary for the proof that follows, Bayes' rule can be used to compute $p(c = k|A = i)$ as follows:

$$p(c = k|A = i) = \frac{p(A = i|c = k)\lambda_k}{p(A = i)} \tag{2.6}$$

or

$$p(c = k|A = i) = \frac{p(A = i|c = k)\lambda_k}{p(A = i|c = 1)\lambda_1 + p(A = i|c = 2)\lambda_2}. \tag{2.7}$$

Since $E_F$ is the minimum over all pure state decompositions of $\varrho_{BA}^{(x)}$, we have

$$\bar{E}_F\left(\rho_{BA}^{(x)}\right) \leq \pi_{A,0} p(c=1|A=0) E(\rho_{BA|0,1}) + \pi_{A,0} p(c=2|A=0) E(\rho_{BA|0,2}) \\ + \pi_{A,1} p(c=1|A=1) E(\rho_{BA|1,1}) + \pi_{A,1} p(c=2|A=1) E(\rho_{BA|1,2}). \tag{2.8}$$

Rearranging terms and forming joint probabilities to the above

$$\bar{E}_F\left(\rho_{BA}^{(x)}\right) \leq p(c=1 \cap A=0) E(\rho_{BA|0,1}) + p(c=1 \cap A=1) E(\rho_{BA|1,1}) + \\ p(c=2 \cap A=0) E(\rho_{BA|0,2}) + p(c=2 \cap A=1) E(\rho_{BA|1,2}), \tag{2.9}$$

$$= \lambda_1 \left(p(A=0|c=1) E(\rho_{BA|0,1}) + p(A=1|c=1) E(\rho_{BA|1,1})\right) + \\ \lambda_2 \left(p(A=0|c=2) E(\rho_{BA|0,2}) + p(A=1|c=2) E(\rho_{BA|1,2})\right). \tag{2.10}$$

Now, from Eq. (2.3) the equation above is given by

$$= \lambda_1 \bar{E}(|\phi_1\rangle) + \lambda_2 \bar{E}(|\phi_2\rangle), \tag{2.11}$$

and from Eq. (2.4) applied to both $\bar{E}(|\phi_1\rangle)$ and $\bar{E}(|\phi_2\rangle)$:

$$\bar{E}_F\left(\rho_{BA}^{(x)}\right) \leq \lambda_1(1-D) + \lambda_2(1-D). \tag{2.12}$$

Therefore $\bar{E}_F\left(\rho_{BA}^{(x)}\right) \leq 1 - D$ or $\bar{E}_F\left(\rho_{BA}^{(x)}\right) + D \leq 1$.

## 2.2. Two-particle Measurement

In Appendix 2.1, Alice and Bob share a mixed state and Alice makes a measurement on Qubit $A$. In this section on two-particle measurement, Alice and Bob similarly initially share a mixed state and a measurement is then made on Alice's qubit resulting in an average amount of entanglement. However, additionally a second measurement is then made on Bob's qubit in order to maximize the average information gain $G$. As in Appendix 2 we again suppose that unbeknownst to Bob and Alice, Charles can either insert the improper mixed state $\rho_{BA}^{(x)} = \sum_{i=1}^{N} \lambda_k |\phi_k\rangle\langle\phi_k|$ or perform a pure-state pre-selection for which $|\phi_k\rangle$ is chosen with probability $\lambda_k$. Additionally, if Charles performs a pure-state pre-selection he has the option of providing Bob the result of the pre-selection.

The average entanglement of formation that results when Charles inserts the improper mixed state $\rho_{BA}^{(x)}$ between Alice's and Bob's qubits after Alice's measurement is identical to that which was considered in the prior subsection. Hence, we know

$$\bar{E}_F\left(\rho_{BA}^{(x)}\right) \leq \pi_{A,0} p(c=1|A=0) E(\rho_{BA|0,1}) + \pi_{A,0} p(c=2|A=0) E(\rho_{BA|0,2}) \\ + \pi_{A,1} p(c=1|A=1) E(\rho_{BA|1,1}) + \pi_{A,1} p(c=2|A=1) E(\rho_{BA|1,2}). \tag{2.13}$$

or

$$\bar{E}_F\left(\rho_{BA}^{(x)}\right) \leq p(c=1)\bar{E}(|\phi_1\rangle) + p(c=2)\bar{E}(|\phi_2\rangle). \tag{2.14}$$

We desire to prove

$$\bar{E}_F\left(\rho_{BA}^{(x)}\right) + \bar{G}(\rho_{BA}^{(x)}) \leq 1, \tag{2.15}$$

where the gain $G_i \equiv 1 - 2P_{e|A=i}$, $P_{e|A=i}$ is the probability of error given Alice obtains result $i \in \{0,1\}$, and the average gain is found as

$$\bar{G}\left(\rho_{BA}^{(x)}\right) = \pi_{A,0} G_0\left(\rho_{BA}^{(x)}\right) + \pi_{A,1} G_1\left(\rho_{BA}^{(x)}\right) \tag{2.16}$$

Now, $\bar{G}\left(\rho_{BA}^{(x)}\right)$ is the minimum average gain over all detection schemes given that the initial state is $\rho_{BA}^{(x)}$. Consider that Charles performs a pure-state preselection and additionally tells Bob which state he preselected. Bob then can compute two optimal decision techniques: one given Charles selects $k = 1$ and another when Charles selects $k = 2$. Suppose that when Charles selects the pure state $k$, the information gain of Bob (averaged over Alice's outcomes) is denoted by $\bar{G}_{C,k}$ and the information gain further averaged over Charles's possible choices $k \in \{1,2\}$ denoted by $\bar{G}_C$. Hence $\bar{G}_C = p(c=1)\bar{G}_{C,1} + p(c=2)\bar{G}_{C,2}$. Note that Charles has the option of continuing to employ the technique used without knowledge of Bob's pre-selection thereby ignoring Charles's information, but such a technique must be the same or suboptimal in terms of maximizing the gain when compared with the optimal technique used with knowledge of Charles's preselection. Therefore $\bar{G}\left(\rho_{BA}^{(x)}\right) \leq \bar{G}_C$. Furthermore,

$$\bar{G}\left(\rho_{BA}^{(x)}\right) \leq p(k=1)\bar{G}_{C,1} + p(k=2)\bar{G}_{C,2}. \tag{2.17}$$

Hence, we have

$$\bar{E}_F\left(\rho_{BA}^{(x)}\right) + \bar{G}\left(\rho_{BA}^{(x)}\right)$$
$$\leq p(c=1)\bar{E}(|\phi_1\rangle) + p(c=2)\bar{E}(|\phi_2\rangle) + p(c=1)\bar{G}_{C,1} + p(k=2)\bar{G}_{C,2}. \quad (2.18)$$

$$= p(c=1)\left(\bar{E}(|\phi_1\rangle) + \bar{G}_{C,1}\right) + p(c=2)\left(\bar{E}(|\phi_2\rangle) + \bar{G}_{C,2}\right). \quad (2.19)$$

Now, we know from Sec. 3.2 that for any initial pure state that $\bar{E} + \bar{G} \leq 1$, where the averaging in $\bar{E}$ and $\bar{G}$ is over Alice's outcomes $i \in \{0,1\}$. From above, both the expressions $\bar{E}(|\phi_1\rangle) + \bar{G}_{C,1} \leq 1$ and $\bar{E}(|\phi_2\rangle) + \bar{G}_{C,2} \leq 1$ follow. Therefore,

$$\bar{E}_F\left(\varrho_{BA}^{(x)}\right) + \bar{G}\left(\rho_{BA}^{(x)}\right) \leq 1, \quad (2.20)$$

and the result in Sec. 3.2 is also established for arbitrary initial mixed states.

## Appendix 3. Non-Hermitian Operations

Note that in forming a POVM, the POVM elements $E_i = M_i^{(A)\dagger} M_i^{(A)}$ are always Hermitian, however the $M_i^{(A)}$ need not be Hermitian. In this subsection, it is shown that given arbitrary positive-semidefinite elements $E_i$ and a non-Hermitian $M_i^{(A)}$ implementating $E_i = M_i^{(A)\dagger} M_i^{(A)}$, there exists a related Hermitian implementation $\widehat{M}_i^{(A)}$ for which 1) $E_i = \widehat{M}_i^{(A)\dagger} \widehat{M}_i^{(A)}$, 2) both of Alice's possible outcomes $i = 0,1$ are the same using either $M_i^{(A)}$ or $\widehat{M}_i^{(A)}$, and 3) the matrix that Bob obtains $\rho_{B,i}$ $i = 0,1$ are also identical using either $M_i^{(A)}$ or $\widehat{M}_i^{(A)}$.

Assume $M_i^{(A)}$ is non-Hermitian. There exists a polar decomposition $M_i^{(A)} = PU$ where $P$ is a positive-semidefinite matrix and $U$ is unitary. Now consider the following Hermitian implementation of Alice's measurement $\widehat{M}_i^{(A)} = U^\dagger P U$. Note that the second unitary $U^\dagger$ does not affect the probability of Alice obtaining outcomes $i = 0,1$ since $\hat{\pi}_{A,i} = \text{Tr}[\rho \widehat{M}_i^{(A)\dagger} \widehat{M}_i^{(A)}] = \text{Tr}[\rho P^2] = \text{Tr}[\rho M_i^{(A)\dagger} M_i^{(A)}] = \pi_{A,i}$. Also, the density matrix $\rho_{B,i}$, $i = 0,1$ that Bob obtains is independent of whether or not the second unitary $U'$ is applied. This can be seen as follows:

$$\rho_{B,i} = \frac{\text{Tr}_A[(I \otimes M_i^{(A)})\rho_0(I \otimes M_i^{(A)})^\dagger]}{\pi_{A,i.}} = \frac{\text{Tr}_A[(I \otimes PU)\rho_0(I \otimes PU)^\dagger]}{\pi_{A,i}} \quad (3.1)$$

where $\rho_0$ is the initial density matrix of Alice-Bob. Now, $\rho_{B,i}$ is invariant under local unitary operations on Alice's qubit, i.e. given any Alice-Bob density matrix $\rho$, $\rho_{B,i} = \text{Tr}_A[\rho]$ iff (if and only if) $\rho_{B,i} = \text{Tr}_A[(I\otimes U^\dagger)\rho(I\otimes U)]$. Hence from Eq. (3.1) we then have

$$\rho_{B,i} = \frac{\text{Tr}_A[(I\otimes U^\dagger)(I\otimes PU)\rho_0(I\otimes PU)^\dagger(I\otimes U)]}{\pi_{A,i}}$$

$$= \frac{\text{Tr}_A[(I\otimes U^\dagger PU)\rho_0(I\otimes U^\dagger P^\dagger U)]}{\hat{\pi}_{A,i}}$$

$$= \frac{\text{Tr}_A[(I\otimes \widehat{M}_i^{(A)})\rho_0(I\otimes \widehat{M}_i^{(A)\dagger})]}{\hat{\pi}_{A,i}} \tag{3.2}$$

$$= \hat{\rho}_{B,i}.$$

Hence if Alice utilizes a non-Hermitian implementation of a POVM with $\sum E_i = I$, $M_i^{(A)\dagger}M_i^{(A)} = E_i$, then there exists a corresponding positive semi-definite implementation $\widehat{M}_i^{(A)}$ of the same POVM. We have already seen that any such positive semidefinite implementation $\widehat{M}_i^{(A)}$ can be expressed in the form given of Eq. (1.1) and for which it has already been proven that $\bar{E} + \bar{G} \leq 1$ for such $\widehat{M}_i^{(A)}$. Since the non-Hermitian form has both the same probabilities for Alice's outcomes and as well Bob's density matrices that correspond to Alice's outcomes, with both the average entanglement $\bar{E}$ and the average gain $\bar{G}$ then $\bar{E} + \bar{G} \leq 1$ continues to hold when such non-Hermitian $M_i^{(A)}$ implementations of a POVM are employed.

## Appendix 4. Treatment of von Neumann Measurement

In the main text of the paper, Alice was assumed to make a measurement on her qubit via a weak measurement. Additionally, the justification of the use of $F$ in [2] is summarized here for convenience. A von Neumann-type measurement [3] on a spin-1/2 particle is considered that is initially in the superposition state $|\psi_S\rangle = \sqrt{\beta}\,|\uparrow\rangle + \sqrt{1-\beta}\,|\downarrow\rangle$ with $0 \leq \beta \leq 1$, $|\uparrow\rangle$ and $|\downarrow\rangle$ denote the spin-up and spin-down state respectively and let $|\psi_P\rangle$ denote the state of the pointer which is initially set to $|\psi_{P,0}\rangle$. In the von Neumann scheme, the system and the measurement device are allowed to interact via Schrödinger's equation and become entangled so that the system and device pointer evolve to the final state. After the interaction, the spin state becomes in the von Neumann scheme entangled with the pointer:

$$|\psi_S\rangle \otimes |\psi_{P,0}\rangle \rightarrow |\psi_{S,P}\rangle \equiv \sqrt{\beta}\,|\uparrow\rangle\otimes|\psi_{P,\uparrow}\rangle + \sqrt{1-\beta}\,|\downarrow\rangle\otimes|\psi_{P,\downarrow}\rangle. \tag{4.1}$$

where $|\psi_{P,\uparrow}\rangle$ denotes the final pointer state when spin-up is the system state and $|\psi_{P,\downarrow}\rangle$ when spin-down is the system state, and $|\psi_{S,P}\rangle$ is the system pointer entangled state. For initial system states $\rho_S$, the average final density operator $\bar{\rho}_{S,F}$ of the system without further projective measurement can be found by tracing out the pointer in Eq. (4.1) giving:

$$\bar{\rho}_{S,F} = F\rho_S + (1-F)(P_\uparrow \rho_S P_\uparrow + P_\downarrow \rho_S P_\downarrow) \tag{4.2}$$

where $P_\uparrow = |\uparrow\rangle\langle\uparrow|$, $P_\downarrow = |\downarrow\rangle\langle\downarrow|$ and $F = \langle\psi_{P,\downarrow}|\psi_{P,\uparrow}\rangle$. $F$ characterizes the overlap between the two pointer readings that correspond to spin-up and spin-down. The measurement disturbance $D = 1 - F$ is defined as the overlap between the two pointer readings $F = \langle\psi_{P,\downarrow}|\psi_{P,\uparrow}\rangle$ [2, 4, 5]. If the overlap $F$ is large, then little is learned of the actual spin by the weak measurement. If $F$ is small, then the strength of the measurement is high. One can see from the analysis in [6] that the overlap $F = \langle\psi_{P,\downarrow}|\psi_{P,\uparrow}\rangle$ is still a valid measure of measurement weakness and strength for mixed initial system states $\rho_S$ for our applications in Sec. 3.1 and Appendix 2.

Note that for weak measurement, $|\psi_{P,\uparrow}\rangle$ and $|\psi_{P,\downarrow}\rangle$ are generally non-orthogonal; still the measurement device ultimately registers distinct results of +1 and -1 corresponding to the spin states $\{|\uparrow\rangle, |\downarrow\rangle\}$ which must be orthogonal. Define these measurement device states as $|\phi_+\rangle$ and $|\phi_-\rangle$ corresponding to measuring spin $|\uparrow\rangle$ and $|\downarrow\rangle$ respectively. A projective measurement then occurs on the pointer state so that $|\psi_P\rangle$ is projected in the basis $\{|\phi_+\rangle, |\phi_-\rangle\}$. A question is what are the orthogonal states $\{|\phi_+\rangle, |\phi_-\rangle\}$ that are projected? Both [2, 4] model the measurement pointer using a single qubit [7] via $|\psi_{P,\uparrow}\rangle = \cos\theta_Z |0\rangle + \sin\theta_Z |1\rangle$ and $|\psi_{P,\downarrow}\rangle = \sin\theta_Z |0\rangle + \cos\theta_Z |1\rangle$. The states $|\phi_+\rangle = |0\rangle$, $|\phi_-\rangle = |1\rangle$ were then utilized by Zhu as the projection states in [4]. Note that the choice of $|\phi_+\rangle = |0\rangle$, $|\phi_-\rangle = |1\rangle$ provide orthogonal measurement device states that correspond to the spin states $\{|\uparrow\rangle, |\downarrow\rangle\}$ in the limit of strong measurement and continue to provide correlation to the spin states $\{|\uparrow\rangle, |\downarrow\rangle\}$ in the case of weak measurement. After a measurement occurs on the pointer states using the basis $\{|\phi_+\rangle, |\phi_-\rangle\} = \{|0\rangle, |1\rangle\}$ the von Neumann measurement scheme has also been shown to be completely equivalent to a weak measurement on the original qubit using the measurement operators $\{M_+, M_-\}$ in Ref. [4] i.e. without having to first unitarily entangle the system and measuring device. This latter equivalence will be used to represent measurement and can be readily computed from a measurement quality factor $F = \langle\psi_{P,\downarrow}|\psi_{P,\uparrow}\rangle$ as $2\sin\theta_Z \cos\theta_Z = \sin 2\theta_Z$. Assuming $0 \leq \theta_Z \leq \frac{\pi}{4}$, note from Sec. 2.3 that $\theta_Z = \text{acos} \frac{\Lambda+1}{\sqrt{1+2\Lambda(1+\Lambda)}}$ or $\cos\theta_Z = \frac{\Lambda+1}{\sqrt{1+2\Lambda(1+\Lambda)}}$, from which follows $\sin\theta_Z = \frac{\Lambda}{\sqrt{1+2\Lambda(1+\Lambda)}}$ and $F = \frac{2\Lambda(1+\Lambda)}{1+2\Lambda(1+\Lambda)}$.

Note that the class of positive-semidefinite matrices $\{M_+, M_-\}$ in Ref. [4] can be directly related to our description of weak measurement in Eqn. (5) as follows:

$$M_\pm = \cos\theta_Z |k_\pm\rangle\langle k_\pm| + \sin\theta_Z |k_\mp\rangle\langle k_\mp| \tag{4.3}$$

with $0 \leq \theta_Z \leq \frac{\pi}{4}$. By setting, $|k_-\rangle = -\sqrt{1-a_Z}|0\rangle + \sqrt{a_Z}\,e^{-i\phi_Z}|1\rangle$, $|k_+\rangle = \sqrt{a_Z}|0\rangle + \sqrt{1-a_Z}\,e^{-i\phi_Z}|1\rangle$, $a_Z = \cos^2\theta$, $\theta_Z = \mathrm{acos}\frac{\Lambda+1}{\sqrt{1+2\Lambda(1+\Lambda)}}$, $\phi_Z = \phi$, we find that $\langle k_-|k_+\rangle = 0$, $M_0^{(A)} = M_+$, $M_1^{(A)} = M_-$.

Supplementary Material References